\documentclass{svproc}
\usepackage{url}

\usepackage{graphicx}
\usepackage{caption}
\usepackage{subcaption}
\usepackage{xurl}
\usepackage{stackengine,graphicx,xcolor}
\usepackage{tikz}

\begin{document}
\mainmatter              
\title{A Study on the Resource Utilization and User Behavior on Titan Supercomputer}
\titlerunning{SMC Data Challenge 4}  
\author{Sergio Iserte}
\authorrunning{S. Iserte} 
\institute{Universitat Jaume I, Spain.\\
\email{siserte@uji.es}}

\maketitle              

\begin{abstract}
Understanding HPC facilities users' behaviors and how computational resources are requested and utilized is not only crucial for the cluster productivity but also essential for designing and constructing future exascale HPC systems.

This paper tackles Challenge 4, `Analyzing Resource Utilization and User Behavior on Titan Supercomputer', of the 2021 Smoky Mountains Conference Data Challenge. Specifically, we dig deeper inside the records of Titan to discover patterns and extract relationships.

This paper explores the workload distribution and usage patterns from resource manager system logs, GPU traces, and scientific areas information collected from the Titan supercomputer. Furthermore, we want to know how resource utilization and user behaviors change over time.

Using data science methods, such as correlations, clustering, or neural networks, our findings allow us to investigate how projects, jobs, nodes, GPUs and memory are related. We provide insights about seasonality usage of resources and a predictive model for forecasting utilization of Titan Supercomputer. In addition, the described methodology can be easily adopted in other HPC clusters.

\end{abstract}

\section{Introduction}
High-performance computing (HPC) systems are facilities composed of large amounts of computational resources interconnected.
This architecture allows computers to collaborate in the solution of a particular problem.
Moreover, these systems are not expected to have a single user.
Instead, HPC facilities are shared among hundreds or thousands of users coming from very different areas of knowledge.

Next-generation of HPC clusters is expected to reach exascale performance, which implicitly implies an unprecedented growth in the number of computational resources. Probabilistically, the more resources the higher rate of hardware failures. 
Moreover, increasing the pool of resources enables the submission of more jobs and larger requests.

For these reasons, understanding how HPC clusters are utilized is not only crucial to detect productivity issues but also to improve the design of both job scheduling policies and future HPC systems~\cite{9355229}.

Each year, Oak Ridge National Laboratory (ORNL) and the Smoky Mountains Computational Sciences and Engineering Conference (SMC) publishes a series of data science challenges, known as the SMC Data Challenge. 
In the 2021 edition, Challenge 4 \textit{Analyzing Resource Utilization and User Behavior on Titan Supercomputer}~\cite{Dash2021} put the spotlight on how a particular HPC system has been utilized.
In this regard, the challenge presents two datasets with data from the year 2015 to 2019 of the Titan supercomputer, which remained in the top 10 of the TOP500 list for a long time~\cite{top500}. 
Titan was a Cray XK7 system composed of 18,688 nodes AMD Opteron 6274 16-core with 32 GB of DDR3 ECC memory and each one equipped with an Nvidia Tesla K20X GPU with 6GB of GDDR5 ECC memory~\cite{OakRidgeNationalLaboratory2012}.
The first dataset (\texttt{RUR} dataset) contains the scheduler traces of submitted jobs which bring information about users' requests~\cite{Wang2019}, while the second (\texttt{GPU} dataset) compiles GPU hardware-related failures~\cite{Ostrouchov2020}. 

Resource utilization statistics of submitted jobs on a supercomputer can help us understand how users from various scientific domains use HPC platforms, in turn, design better job scheduling policies. Thanks to the data gathered from schedulers, hardware, and users, we can analyze resources utilization and how the behavior of users may change over time adapting to given circumstances.

Concretely, in this paper, a thorough analysis of Titan's log is performed to provide a better understanding of user behaviors and how resources are used in this facility.
The rest of the paper is structured as follows: Section~\ref{sec:explor} presents a preliminary datasets exploration which helps to understand which information is available, and how to make the most of it. This section tackles tasks 1 and 2 from the proposed challenge.
Section~\ref{sec:series} explores the data over time and studies behaviors in different year seasons and front of system failures. The section also includes a predictive model trained on Titan data.
This section addresses tasks 3 and 4 from the proposed challenge.
Finally, in Section~\ref{sec:concl} most remarkable conclusions are highlighted.

\section{Exploratory Data Analysis}\label{sec:explor}
This section deals with tasks 1 and 2 from the challenge.
On the one hand, an exploratory data analysis on the \texttt{RUR} dataset to summarize data characteristics is performed. We also investigate if there are relationships between client CPU memory usage, GPU memory usage, and the job size (number of compute nodes).
On the other side, using clustering methods we will see if there are similarities in the resource usage patterns among jobs based on the projects to which they belong.

\subsection{Data Preprocessing}\label{subsec:preproc}
The job scheduling dataset (\texttt{RUR}) has the traces of 12,981,186 jobs submitted by a total of 2,372 users.
To begin with, raw data is preprocessed and loaded into a clean dataset. 
For this purpose, from the \texttt{RUR} dataset we leverage the following fields:

\begin{itemize}
    \item \texttt{node\_count}: Number of nodes requested by the job.
    \item \texttt{max\_rss}: Estimation of the maximum resident CPU memory used by an individual compute node through the lifespan of a job run.
    \item \texttt{start\_time}: The timestamp when the job started.
    \item \texttt{end\_time}: The timestamp when the job ended.
    \item \texttt{alps\_exit}: Job status upon completion.
    \item \texttt{command}: The executable path from where the information about the area of knowledge and project of the job can be extracted.
    \item \texttt{gpu\_maxmem}: Maximum GPU memory used by all the nodes of the job.
    \item \texttt{gpu\_summem}: Total GPU memory used by all the nodes of the job.
    \item \texttt{gpu\_secs}: Time spent on GPUs by the job.
\end{itemize}

The preprocessing method will remove incomplete, as well as not relevant, traces in order to reduce noise or outliers, which is translated into a higher quality dataset.
For this purpose, the preprocessing consists of the following actions:
\begin{enumerate}
    \item Filter our jobs that did not run successfully (\texttt{alps\_exit} value different from zero). This involves canceled or failed jobs, among other causes that prevented jobs to be completed with success.
    \item Job duration calculation from the difference between end time and start time.
    \item Dismiss jobs without GPU time assigned. In other words, jobs that did not use GPU acceleration.
    \item Remove jobs shorter than 13 minutes. Short jobs may be understood as tests or debugging jobs. This action can be understood as a noise remover that highlights interesting correlations. Thus, by dropping those jobs the correlations (direct or indirect) following depicted are stronger.
    \item Extract the project and area identifiers from \texttt{command} field.
    \item Filter out incomplete job traces without information about the project or area.
\end{enumerate}

After the sanitation, the clean dataset contains 685,101 rows, in other words, jobs.

\subsection{Data Correlation}\label{subsec:corr}
To study relationships among the target fields, we have opted to visualize them with a heat map that highlights the strengths of their correlations.

Since our clean dataset contains numerical and nominal values the correlation mechanisms have to discriminate between types: numerical and nominal.
Thus, although the project identifier is an integer, it is treated as a nominal field.

On the one hand, we have leveraged Person's R between numerical-numerical (continuous) values.
With Pearson's R we measure the linear correlation between sets. The coefficient ranges from $-1$ to $1$, indicating the positive value a strong correlation, while the negative values an indirect correlation.
On the other hand, between numerical-nominal values, we have used the correlation ratio.
The correlation ratio of two sets is defined as the quotient between their standard deviations.
It ranges from $0$ to $1$, being $1$ the highest correlation, and $0$ no correlation at all.

\begin{figure}[htp]
    \centering
    \includegraphics[trim={0 0 0 0}, clip, width=1\linewidth]{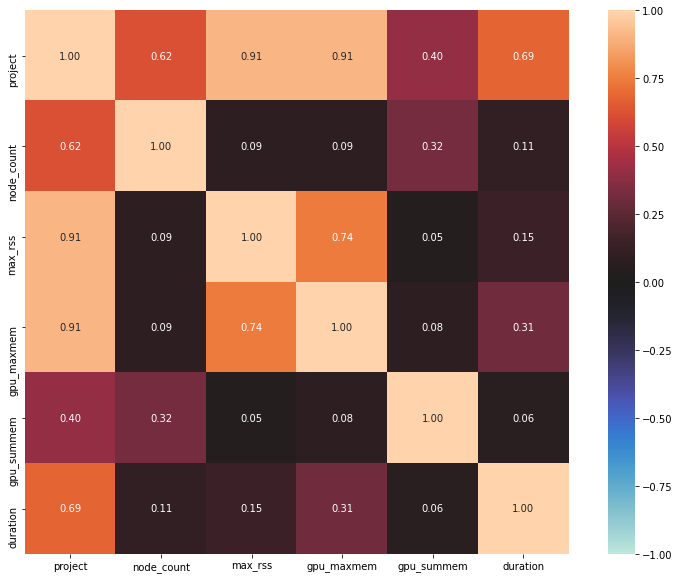}
    \caption{Correlation matrix after the data preprocessing.}
    \label{fig:corr2}
\end{figure}

Figure~\ref{fig:corr2} showcases the correlation matrix with the ratios between tuples of the fields under study.

This study considers a \textit{strong correlation} those tuples with ratio values above 0.5.
In this regard, we can see that the job project is highly correlated to the CPU and GPU memory. Likewise, the project identifier is related to the job duration and the number of nodes assigned to the job.
This can be understood as that long or large jobs are associated with the same projects.

Putting aside the project identifier, the CPU-GPU job memory relationship shows a meaningful direct correlation. Nevertheless, the number of nodes field does not show an important correlation with memory usage.

Since only maximums data of memory were used, the sum of GPU memory is included in the study.
However, the ratios with the number of nodes or the project are still not significant enough to be accepted as significantly correlated.

\subsection{Data Clustering}\label{subsec:clus}
Thanks to the data preprocessing performed in Section~\ref{subsec:preproc}, jobs can be classified by their scientific domain.
In this point, we try to discover if jobs can be clustered by their resource usage and, in the case of the existence of well-defined clusters if those groups present similarities or relations with the area of knowledge assigned to their jobs.
For this purpose, clustering techniques are leveraged.
Particularly, K-means clustering~\cite{Jin2010} is one of the most popular methods for this endeavor. K-means methodology is: given an initial point for each cluster, relocate it to its new nearest center, update the centroid by calculating the mean of the member points, and repeat the process until a convergence criterion is satisfied. This process is done for each cluster centroid.

For this clustering study, the following resource usage metrics seemed the most interesting: \texttt{node\_count}, \texttt{max\_rss}, and \texttt{gpu\_maxmem}.

\begin{figure}[htp]
    \centering
    \includegraphics[trim={0.5cm 0.5cm 0.5cm 0.25cm}, clip, width=1\linewidth]{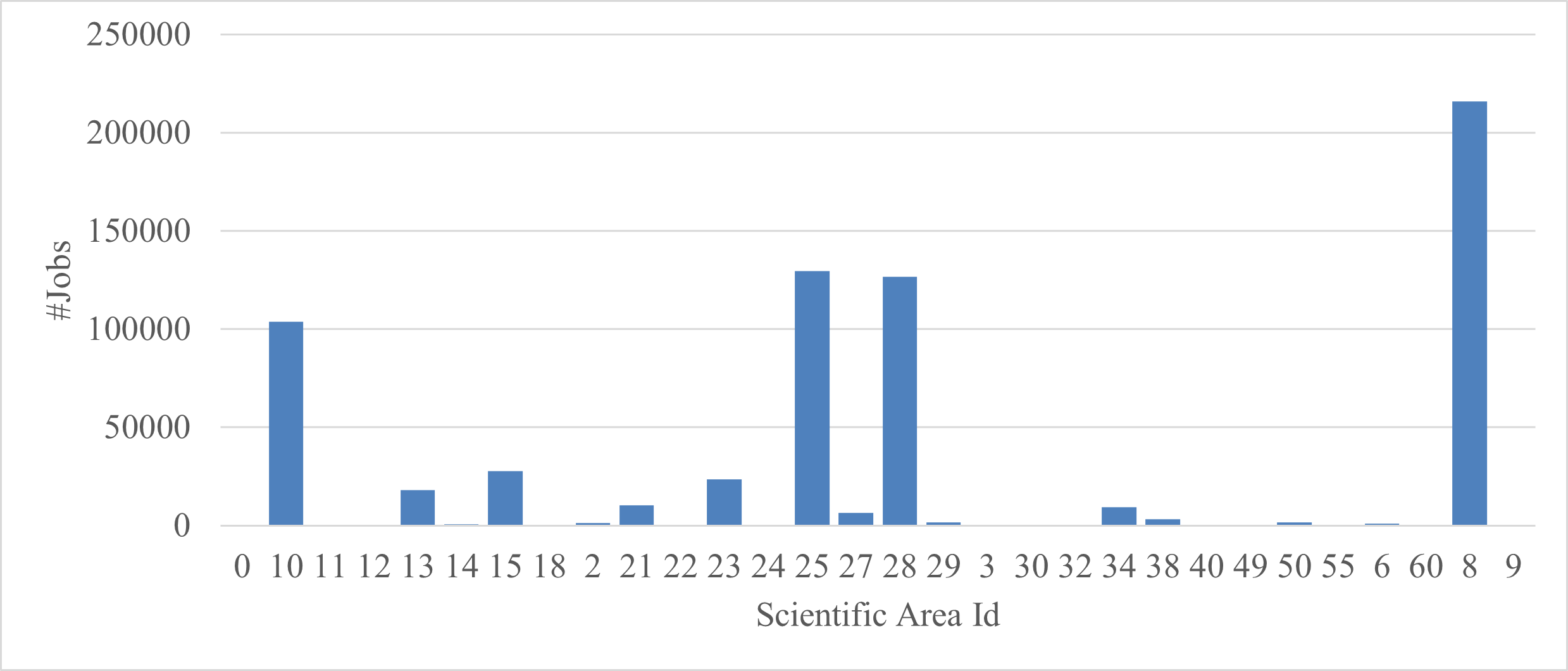}
    \caption{Number of jobs assigned to each area.}
    \label{fig:areas}
\end{figure}

To begin with, the distribution of jobs per domain is analyzed. 
For a total of 30 scientific areas, Figure~\ref{fig:areas} contains a histogram that depicts the number of jobs belonging to each of them.

Among the 30 areas, four of them stand out with a higher number of jobs.
These four predominant areas correspond to \textit{Chemistry} (10), \textit{Lattice Gauge Theory} (25), \textit{Materials Science} (28), and \textit{Biophysics} (8), respectively in the figure from left to right. 

Coincidentally, when using the \textit{elbow method} to get an insight into the number of possible clusters in which the data could be grouped, we obtain the same number of clusters, four.

However, when applying the K-means algorithm to the dataset, the clustering is not conclusive enough since 99.95\% of the samples are clustered together. Leaving the 0.05\% to the three remaining clusters.

In conclusion, the heavy unbalance share of jobs among areas, and the apparent similarities in their usage make clustering not helpful enough to search for relationships.

\section{Time Series Analysis}\label{sec:series}
Challenges 3 and 4 are tackled in this section.
Initially, we study if there is any seasonal impact on resources utilization. Furthermore, \texttt{GPU} dataset with hardware-related issues is aligned to the \texttt{RUR} dataset information, in order to characterize relationships between them.
Finally, research on the time series extracted from the datasets is conducted with the help of a predictive model.

\subsection{Seasonality}\label{subsec:season}
\begin{figure}[htp]
    \centering
    \includegraphics[trim={0 0 0 0}, clip, width=1\linewidth]{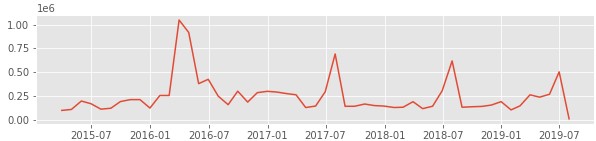}
    \caption{Count of run jobs (y-axis) per month (x-axis).}
    \label{fig:jobs}
\end{figure}

This study aims to detect recurrent patterns through the seasons of the year.
For this purpose, we initially start the study by understanding how many jobs Titan has run in each month.
Figure~\ref{fig:jobs} showcases this metric. 
Although there is no certain pattern along with the time series, last year shows spikes in the number of running jobs during July-August. 

Seasonality is studied with the technique of Season-Trend decomposition using LOESS (STL)~\cite{cleveland1990stl}, grouping data by months.
Furthermore, depending on the analyzed metric, the aggregation method will vary within the mean, maximum, or sum.

\begin{figure}
     \centering
     \begin{subfigure}[b]{1\textwidth}
         \centering
        \includegraphics[trim={40 0 0 0}, clip, width=1\linewidth]{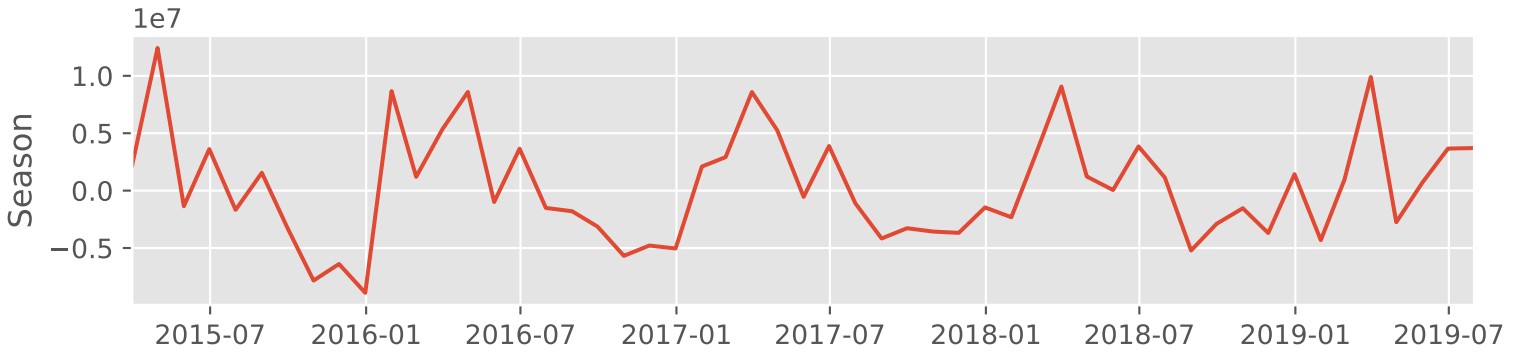}
        \caption{Node count (y-axis).}
        \label{subfig:sum-nodes}
     \end{subfigure}
     \\
     \begin{subfigure}[b]{1\textwidth}
         \centering
        \includegraphics[trim={45 0 0 0}, clip, width=1\linewidth]{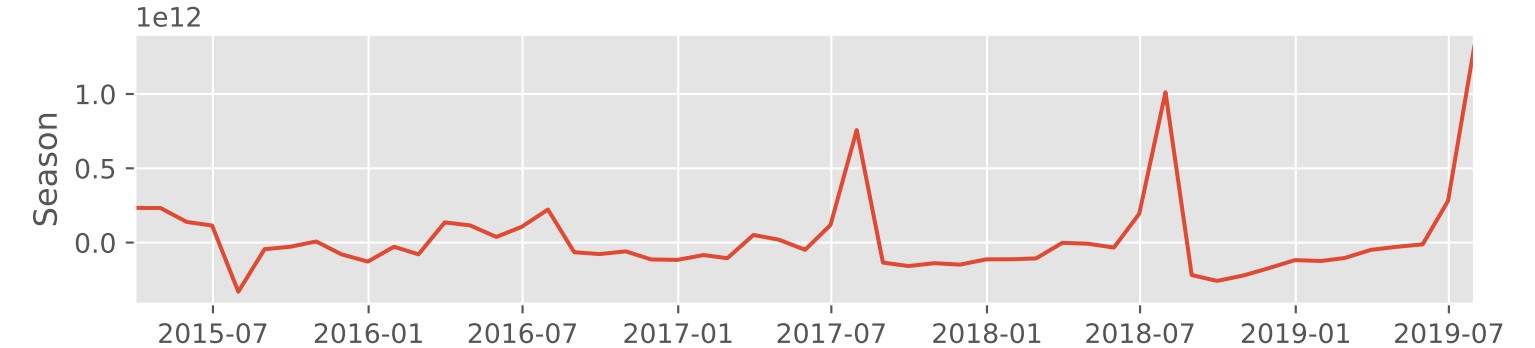}
        \caption{CPU memory in kilobytes (y-axis).}
        \label{subfig:sum-rss}
     \end{subfigure}
     \\
     \begin{subfigure}[b]{1\textwidth}
         \centering
        \includegraphics[trim={40 0 0 0}, clip, width=1\linewidth]{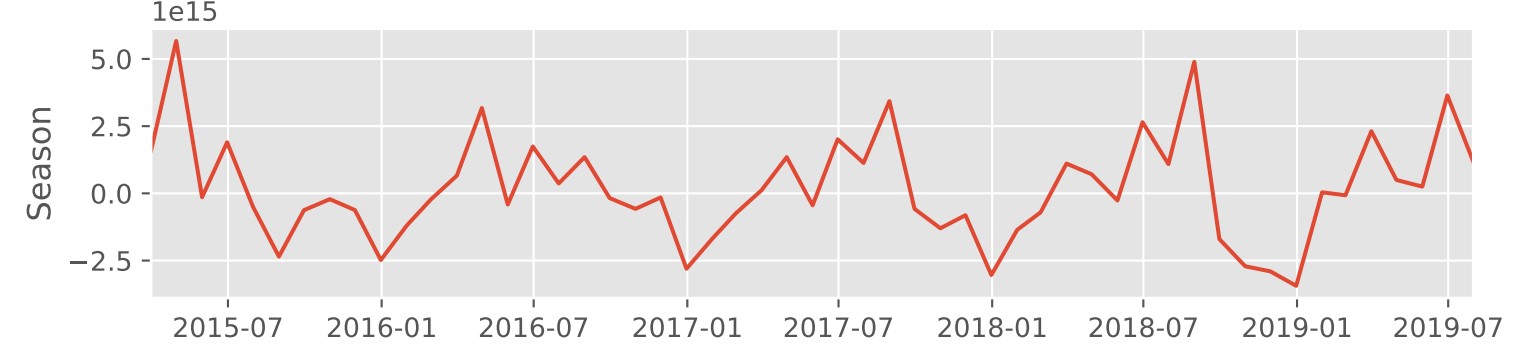}
        \caption{CPU time in seconds (y-axis).}
        \label{subfig:sum-stime}
     \end{subfigure}
        \caption{Seasonality per month aggregated by sum.}
        \label{fig:season-sum}
\end{figure}

Regarding the number of nodes assigned to jobs, we have opted for aggregating with sum. Since the number of nodes in the cluster is known, with this sum we can estimate the aggregated load per month of the cluster. 
However, as it is depicted in Figure~\ref{fig:season-sum}, CPU memory and time also show a seasonal pattern.

In figures~\ref{subfig:sum-nodes} and~\ref{subfig:sum-stime} we can appreciate an increase of nodes assigned to jobs and CPU time, respectively during the first semester of the year. For the second semester, the trend decreases.

The CPU memory (see Figure~\ref{subfig:sum-rss}) presents spikes of usage at the end of the summer season.
Furthermore, CPU time and memory peaks correspond to running job count spikes (see Figure~\ref{fig:jobs}). Likewise, the drop of the spikes coincides with the decrease in the node count.  

\begin{figure}
     \centering
     \begin{subfigure}[b]{1\textwidth}
         \centering
        \includegraphics[trim={70 0 0 0}, clip, width=1\linewidth]{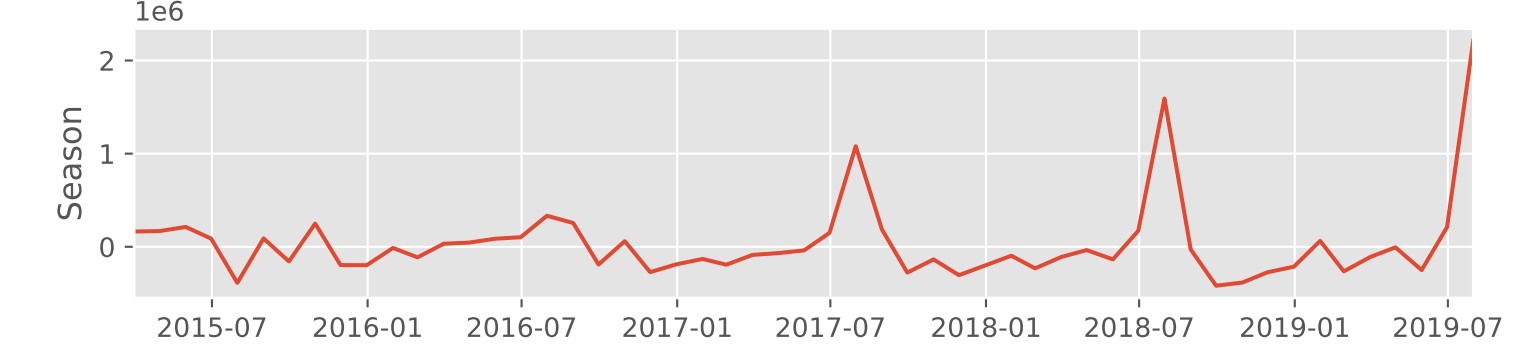}
        \caption{CPU memory in kilobytes (y-axis).}
        \label{subfig:mean-rss}
     \end{subfigure}
     \\
     \begin{subfigure}[b]{1\textwidth}
         \centering
        \includegraphics[trim={40 0 0 0}, clip, width=1\linewidth]{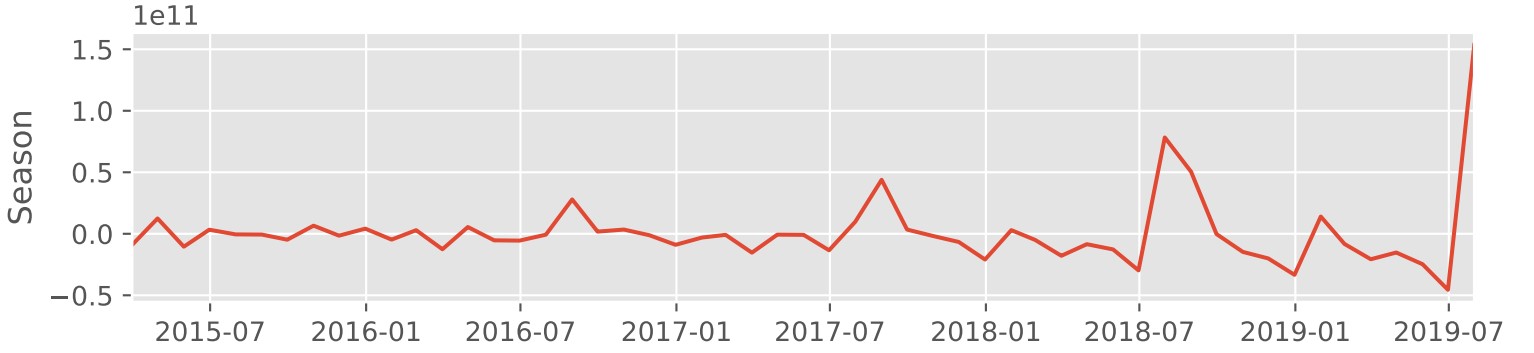}
        \caption{CPU time in seconds (y-axis).}
        \label{subfig:mean-stime}
     \end{subfigure}
        \caption{Seasonality per month aggregated by mean.}
        \label{fig:season-mean}
\end{figure}

In parallel with the sum, mean aggregation (see Figure~\ref{fig:season-mean}) shows the same seasonal patters in CPU time (see Figure~\ref{subfig:mean-stime}) and memory (see Figure~\ref{subfig:mean-rss}).

\begin{figure}
     \centering
     \begin{subfigure}[b]{1\textwidth}
         \centering
        \includegraphics[trim={60 0 0 0}, clip, width=1\linewidth]{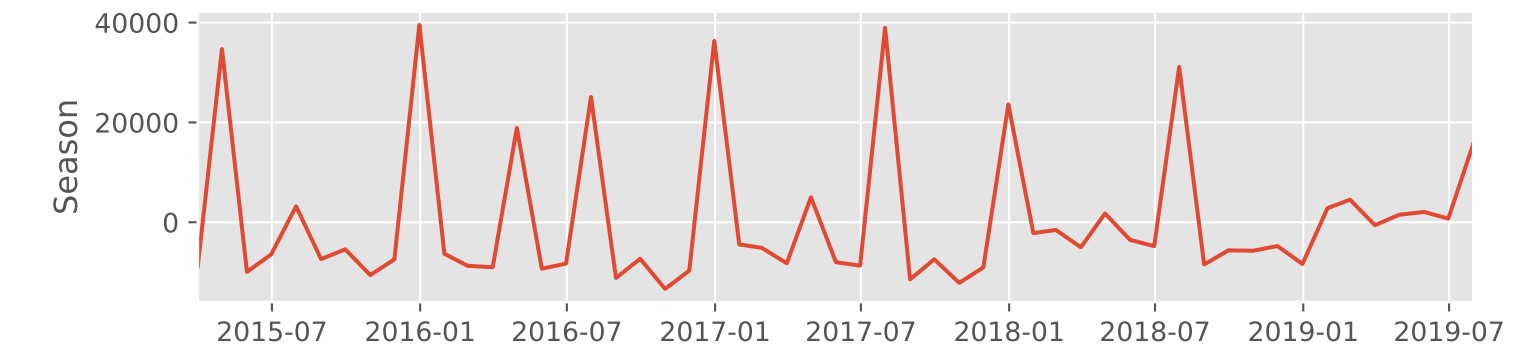}
        \caption{Job duration in seconds (y-axis).}
        \label{subfig:max-time}
     \end{subfigure}
     \\
     \begin{subfigure}[b]{1\textwidth}
         \centering
        \includegraphics[trim={70 0 0 0}, clip, width=1\linewidth]{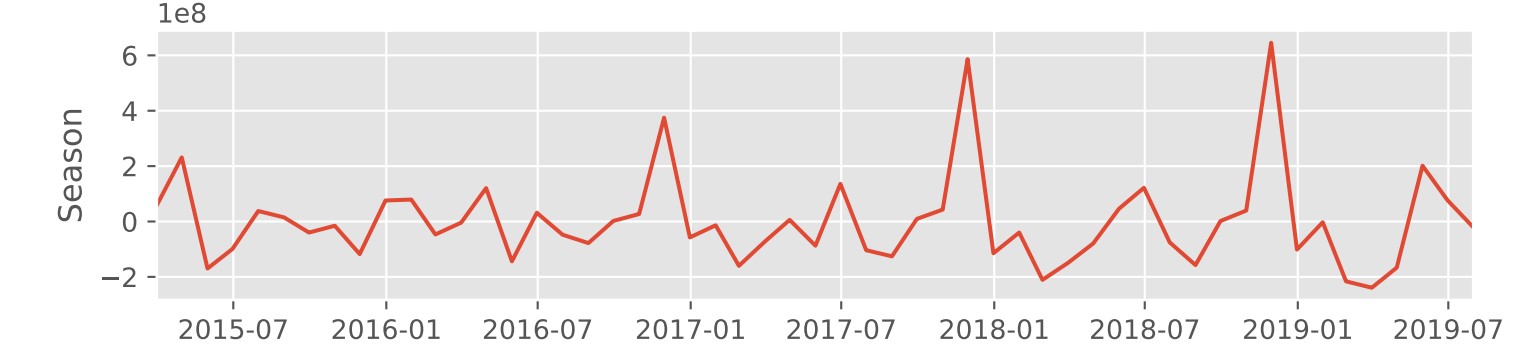}
        \caption{GPU time in seconds (y-axis).}
        \label{subfig:max-gtime}
     \end{subfigure}
        \caption{Seasonality per month aggregated by maximum.}
        \label{fig:season-max}
\end{figure}

When grouping data by months aggregated with maximum values, we find the seasonal patterns showcased in Figure~\ref{fig:season-max}.
Figure~\ref{subfig:max-time} illustrates that very long jobs are usually executed in August and Christmas, probably corresponding to vacation periods, where the job queue is less overloaded.

Figure~\ref{subfig:max-gtime} discovers that long GPU-enabled jobs tend to be executed at the end and in the middle of the year.

Notice that depicted drops in running jobs count, assigned nodes, memory usage, or computation time, may correspond to maintenance shutoffs or power outages.

\subsection{GPU hardware-related issues}
The second dataset available for the challenge is the \texttt{GPU} dataset which logs GPU hardware-related issues on Titan.
From this dataset, we are capable of reconstructing the GPU availability during the records.
For this purpose, fields \texttt{SN}, \texttt{insert} and \texttt{remove}, corresponding to the serial number of a GPU, and times when it was inserted/removed into/from the cluster, respectively, are studied.
Besides, the \texttt{GPU} dataset presented a time-lagged relationship with the \texttt{RUR} dataset, corrected during the preprocessing stage, particularly, removing incomplete data and aligning dates (see Section~\ref{subsec:preproc}).

\begin{figure}[htp]
    \centering
    \includegraphics[trim={0 0 0 0}, clip, width=\linewidth]{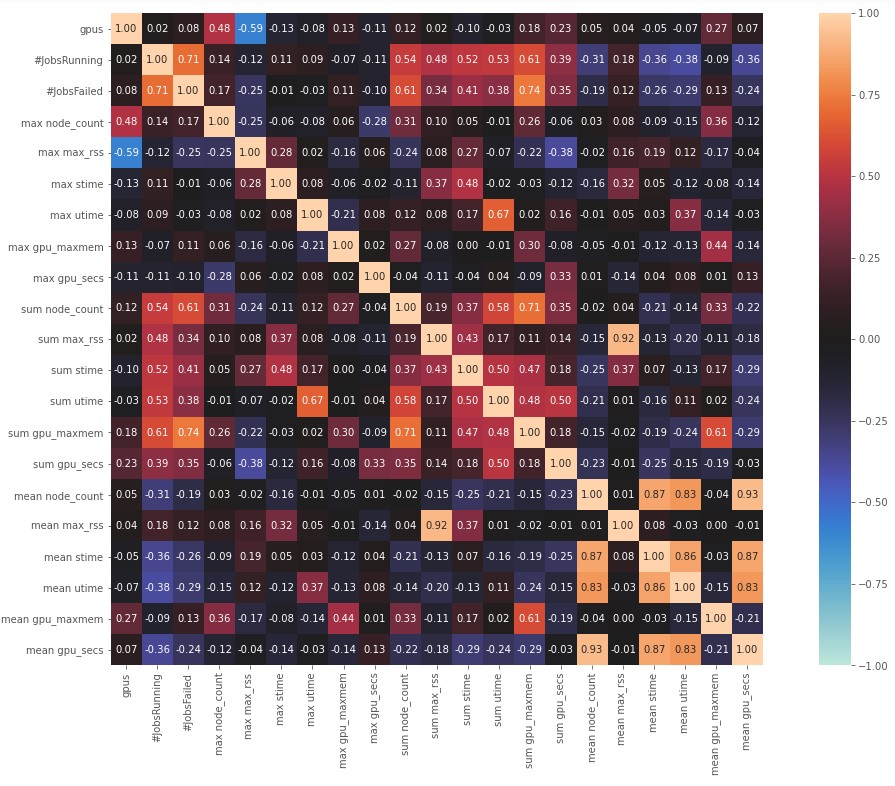}
    \caption{Correlation matrix for aggregated data.}
    \label{fig:corr-task4}
\end{figure}

In this analysis, we study the effect of the changes in the number of GPUs in the workload.
In this regard, initially, both datasets were timely aligned.
Then, data were aggregated in periods of months from which extracting metrics such as sums, maximums, and means.
Figure~\ref{fig:corr-task4} represents the correlation matrix of the aggregated metrics introduced in Section~\ref{sec:series} for the fields described in Section~\ref{sec:explor}. This matrix also includes the count of maximum available GPUs, the count of executed jobs, and the count of failed jobs per month. 

Regarding the GPUs availability, we appreciate a relation with the count of maximum nodes available per month, probably because a node shutdown (or breakdown) implies relocation of its GPUs.
Curiously, this matrix also reveals an indirect relation between the number of GPUs and the maximum memory allocated in the nodes, which could be understood as the more GPUs available the lower the host memory utilized.
In other words, more GPU-bound jobs are executed, as the correlation value between \texttt{\#jobsRunning} and the sum of the GPU memory maximums (\texttt{sum gpu\_maxmem}) indicates.

\begin{figure}[htp]
    \centering
    \includegraphics[trim={0 0 0 0}, clip, width=\linewidth]{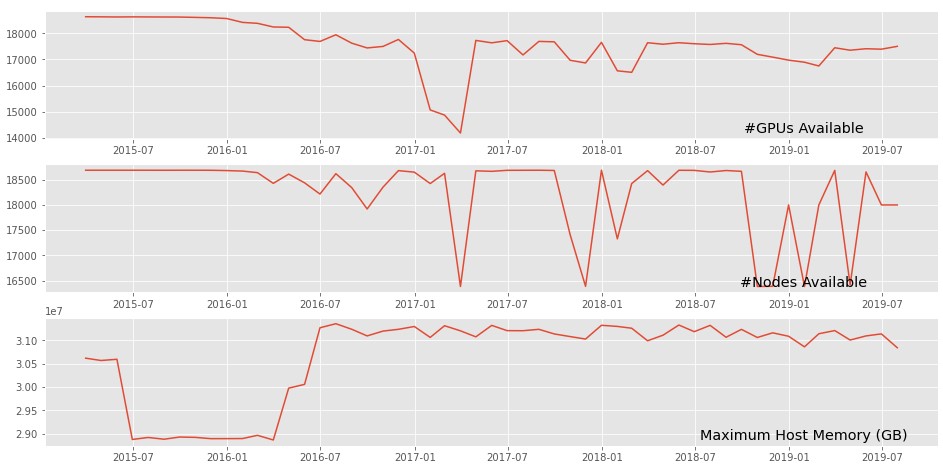}
    \caption{Evolution in time of the resources.}
    \label{fig:evo-task4}
\end{figure}

In order to understand better that statement, Figure~\ref{fig:evo-task4} timely depicts those metrics.
While the relation between GPUs and nodes can be clear, the indirect relation between GPUs and host memory (\texttt{max\_max\_rss}) is not easy to detect. However, it could be determined by the period of late 2015 and early 2016, when the memory records showed a lower utilization.

Furthermore, Figure~\ref{fig:corr-task4} also confirms relationships studied in Section~\ref{sec:explor} thanks to the monthly aggregations.

\subsection{Predictive Model}\label{subsec:forecast}
Given a month's data we can predict the next seven days' values of the four following features: usage of CPU memory (\texttt{max\_rss}), CPU system (\texttt{stime})and user time (\texttt{utime}), and GPU time (\texttt{gpu\_secs}).
For this purpose, we have designed a predictive model based on a recurrent neural network (RNN). Particularly, it is based on the Long Short Term Memory (LSTM) implementation of the RNN.

\begin{figure}[htp]
    \centering
    \includegraphics[trim={0 0 0 0}, clip, width=0.6\linewidth]{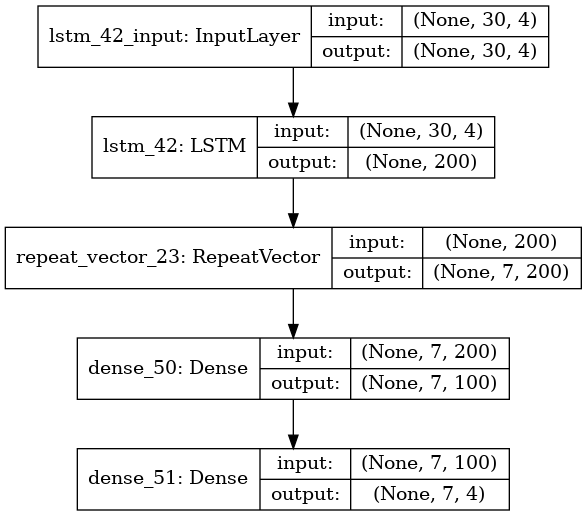}
    \caption{Neural network architecture.}
    \label{fig:nn}
\end{figure}

Figure~\ref{fig:nn} represents the network architecture. 
It expects the data of the four features for 30 days. And it returns the features for the seven following days. 
The model is composed of an LSTM layer with 200 neurons that feeds four parallel double dense layers, one for each feature. This behavior is achieved thanks to repeating the output vector from the recurrent layer (LSTM) with the ``RepeatVector'' layer. The result of this operation is four predictions, one for each of the selected features.
The activation function used for non-linearity is the \textit{relu} which stands for ``Rectified Linear Unit''.

The presented model is compiled with the \textit{Adam} optimizer~\cite{Kingma2015} to update weights and biases within the network, and relies on the mean square error loss function:
\begin{equation}
MSE = \frac{1}{n}\sum_{i=1}^{n}{(y_i-y'_i)^2},
\end{equation}

After normalizing, preparing the data for supervised learning, shuffling, and make partitions of the data (80\% train and 20\% test), the model is evaluated with a 0.4\% error. In other words, predictions present a loss (in this case the MSE) of 0.4\% concerning the expected values.
The model was trained during 11 epochs with a mini-batch size of 16 elements.

This model has proved to be an interesting predictive approach that could be extended straightforwardly to work with more features from Titan's logs.

\section{Conclusions}\label{sec:concl}
In the context of the 2021 SMC Data Challenge, in this work, we explore Challenge 4, `Analyzing Resource Utilization and User Behavior on Titan Supercomputer', and investigate the relationships among jobs, nodes, and GPUs.

Firstly, we have performed an exploratory data analysis on the \texttt{RUR} dataset to summarize data characteristics on the, a priori, some of its most relevant features for the development of this challenge.
Then, we have studied the seasonality of the data.
For this purpose, we load the original \texttt{RUR} dataset without the previous preprocessing since time series is also dependant on job bad terminations, no GPU jobs, or test and debug jobs, among the rest of logged circumstances.
Nevertheless, we have extended this dataset with the job duration time calculated subtracting the end time from the start time.

As part of this investigation, the provided datasets are systematically characterized and prepared to be analyzed. We have devised a methodology to understand seasonal usage patterns and methods to study the relationships. Finally, we provide open-source scripts for conducting the associated analysis and visualizations.
For this purpose, scripts and instructions can be found in \url{https://github.com/siserte/SMC2021-Challenge4.git}.

From the results of our analysis, we can conclude that the available data help to understand user behaviors and resource utilization, as it has been documented through the paper. However, data depicts two well-differentiated stages in the use of Titan split in mid-2017. For this reason, provided data may be insufficient at some points, particularly if the data is timely aggregated.

All in all, the study shows the strong relationship between the projects and how users of those projects reflect similar usage of GPU and CPU memory. Moreover, projects also provide a well-understanding of node counts and job durations. 

We have also detected seasonality patterns that are periodically repeated in the mid and end of the year. Nevertheless, as usual in these types of studies, more quality data means more accurate results and predictions. For this reason, new HPC systems are expected to count with more reliable monitoring and logging mechanisms. 

The presented work may be easily adopted in other HPC facilities since many of them periodically record their users' actions. 
In fact, some insights described in this paper were applied to Cori Supercomputer of NERSC at Lawrence Berkeley National Laboratory (LBNL)\footnote{\url{https://www.nersc.gov/assets/Uploads/NERSC-2019-Annual-Report-Final.pdf}}.

Future resource management systems of HPC facilities will benefit from these kinds of studies, providing their scheduling policies with the experience of past events that are likely to be reproduced. For instance, since jobs that request large amounts of resources tend to wait for long in the queue, systems may schedule their execution taking into account seasonality low peaks where users traditionally submit fewer jobs.
In addition, as we have also seen, the indirect relationship in the number of available GPUs and the host memory usage may make the schedulers collate GPU-enabled jobs with memory-bound jobs to exploit better the resources.

\section*{Acknowledgements}
S.~Iserte was supported by the postdoctoral fellowship APOSTD/2020/026 from Valencian Region Government and European Social Funds\footnote{\url{https://innova.gva.es/va/web/ciencia/a-programa-i-d-i/-/asset_publisher/jMe1UDRYZMHO/content/iv-subvenciones-para-la-contratacion-de-personal-investigador-en-fase-postdoctor-2}}.
The study on Cori supercomputer was carried out during an internship funded under HiPEAC Collaboration Grant H2020-ICT-2017-779656\footnote{\url{https://cordis.europa.eu/project/id/779656}}.
Finally, the author wants to thank the anonymous reviewers whose suggestions significantly improved the quality of this manuscript.

\bibliographystyle{elsarticle-num}
\bibliography{bibliography}

\end{document}